\documentclass{emulateapj} 
\usepackage{psfig}

\slugcomment{  }

\shorttitle{ACS photometry of the globular cluster B514 }
\shortauthors{Galleti et al.}

\begin{document}

\title{ACS photometry of the remote M31 globular cluster B514. 
\thanks{Based on observations with the NASA/ESA Hubble Space Telescope, 
obtained at the Space Telescope Science Institute (STScI), which is operated by
AURA, Inc., under NASA contract NAS 5-26555.}}


\author{S. Galleti\altaffilmark{2}, L. Federici\altaffilmark{2}, 
M. Bellazzini\altaffilmark{2}, A. Buzzoni\altaffilmark{2} and
F. Fusi Pecci\altaffilmark{2}  
}
\altaffiltext{2}{INAF - Osservatorio Astronomico di Bologna, via Ranzani 1, 
I-40127, Bologna, Italy}


\begin{abstract}
We present deep F606W, F814W  ACS photometry of the recently discovered globular
cluster B514, the outermost known globular in the M31 galaxy. The cluster 
appears quite extended and member stars are unequivocally identified out to
$\sim 200$ pc from the center. The Color
Magnitude Diagram reveals a steep Red Giant Branch (RGB), and a 
Horizontal Branch (HB)
extending blue ward of the instability strip, indicating that B514 is a
classical old metal-poor globular cluster. The RGB locus
and the position of the RGB Bump are both consistent with a metallicity
$[Fe/H]\sim -1.8$, in excellent agreement with spectroscopic estimates.
A preliminary estimate of the integrated absolute V magnitude ($M_V\la -9.1$)
suggests that B514 is among the brightest globulars of M31.
\end{abstract}


\section{Introduction}

Stars and stellar systems orbiting the most external regions of their
parent galaxy are crucial tracers of the mass distribution and the formation
history of the parent galaxy as a whole. For example, it is well known that
substructures related to accretion events are best preserved in the outskirts
of a galactic halo \citep[see, for instance,][and references therein]{bj04}.
In this context globular clusters may serve as excellent tracers of
substructures in the outer region of their parent galaxy.
For instance, \citet{bfi} were able to identify
the accretion signature of the  of the Sagittarius dwarf galaxy among the
globular clusters in the outer halo of the Milky Way.

As large substructures are also identified  in the nearby spiral galaxy M31
\citep{m31nat} it becomes increasingly important to have a complete census of
the cluster population in the  galaxy outskirts - still lacking to date
\citep[see][]{G05,G06} - as well as to collect information about cluster
ages, metal content, kinematics, etc. Several groups are currently working 
on this line \citep{hux04,bat}, including ourselves \citep[see][]{G04,G05,G06}.
In particular, in \citet[][hereafter G05]{G05} we reported on the discovery of
the outermost known cluster of M31, B514 located at a projected distance of
$R_p\simeq 55$ Kpc from the center of M31, not far from the galaxy major axis.
The integrated photometry and spectroscopy indicates that B514 is likely an old
metal-poor globular cluster (G05), however direct imaging, at a spatial
resolution sufficient to resolve the system into stars, is the only way to
ultimately ascertain the nature of a candidate globular in M31 
\citep[see][]{G04,G06}.

Here we present the first results of follow-up observations of B514 taken with
the Wide Field Camera (WFC) of the Advanced Camera for Surveys (ACS) on board of
the Hubble Space Telescope (HST), confirming that the object is a genuine old
and metal poor cluster lying in the remotest region of M31. 

\section{Data Analysis}

Observations have been carried out on June 10, 2006 (Prop. Id. GO
10565, P.I. S. Galleti). We acquired three images with the F606W filter, for a
total exposure time of $t_{exp}=2412$ s, and three F814W images, for a
total exposure time of $t_{exp}=2418$ s. The ACS/WFC camera
is a mosaic of two $4096\times 2048$ px$^2$ CCDs, with a pixel scale of $0.05
\arcsec/$px. The pointing was chosen such as to have the
cluster placed near the center of one of the ACS/WFC CCDs (Chip~2), while Chip~1
is presumed to sample the stellar population in the field surrounding the
cluster. Data reduction has been performed on the individual pre-reduced
images provided by STScI, using the ACS module of DOLPHOT\footnote{See
{\tt http://purcell.as.arizona.edu/dolphot/}} \citep{dol}, a
Point Spread Function -fitting package specifically devoted to the photometry of HST data. The
package identifies the sources and performs the photometry on
individual frames taking also into account all the information about image
cosmetics and cosmic-ray hits, that is attached to the observational material. 
DOLPHOT provides as output the magnitudes and positions of the detected sources, as
well as a number of quality parameters for a suitable sample selection, 
in view of the actual scientific objective
one has in mind. Here we selected all the sources having valid magnitude
measurements in both passbands, global quality flag = 1 (i.e., best measured stars),
{\em crowding} parameter $<0.3$, and $\chi^2<1.5$ if F606W$<24.0$, 
and $\chi^2<2.5$ for brighter stars (see \citep{dol} for details on the
parameters). This selection cleans the sample from the vast majority of spurious
and/or bad measured sources whithout significant loss of information.
The magnitudes have been reported to the VEGAMAG system following the
prescriptions of \citet{sirianni}. For reddening corrections, in the 
following we always adopt
$A_{F606}=2.809E(B-V)$ and $A_{F814}=1.825E(B-V)$, from \citet{sirianni}.
These extinction laws are appropriate for a G2 stars, but for $E(B-V)\le 0.1$
they are accurate to within 0.01 mag for a large range of spectral types 
\cite[from O5 to M0, according to][]{sirianni}.


\begin{figure}
\epsscale{.90}
\plotone{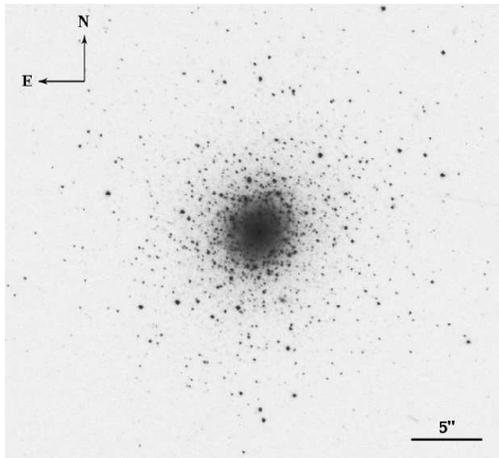}
\caption{B606W image of B514} 
\label{fig1}
\end{figure}

Fig.\ref{fig1} shows a section of the combined (drizzled) F606W image centered on B514.
The exquisite spatial resolution of the ACS/WFC clearly resolves the object into
stars, revealing a beautiful globular cluster with an extended halo. We were
able to obtain useful photometry of individual stars down to a distance of 
$r\simeq 2.5\arcsec$ from the cluster center, that we identified at the position of
the peak in the light profile at $x_{px}=2084$ and $y_{px}=1021$ on Chip 2.
In any case the adopted selection criteria excludes from the adopted sample the
stars located in the most crowded region of the cluster. In particular the
adopted sample contains no stars with $r\le 2.5\arcsec$ and only 44 bright star
(over a total sample of 1562 stars in Chip 1) having $r\le 5\arcsec$.

\section{The Color Magnitude Diagram of B514}

\begin{figure*}[t]
\epsscale{.80}
\plotone{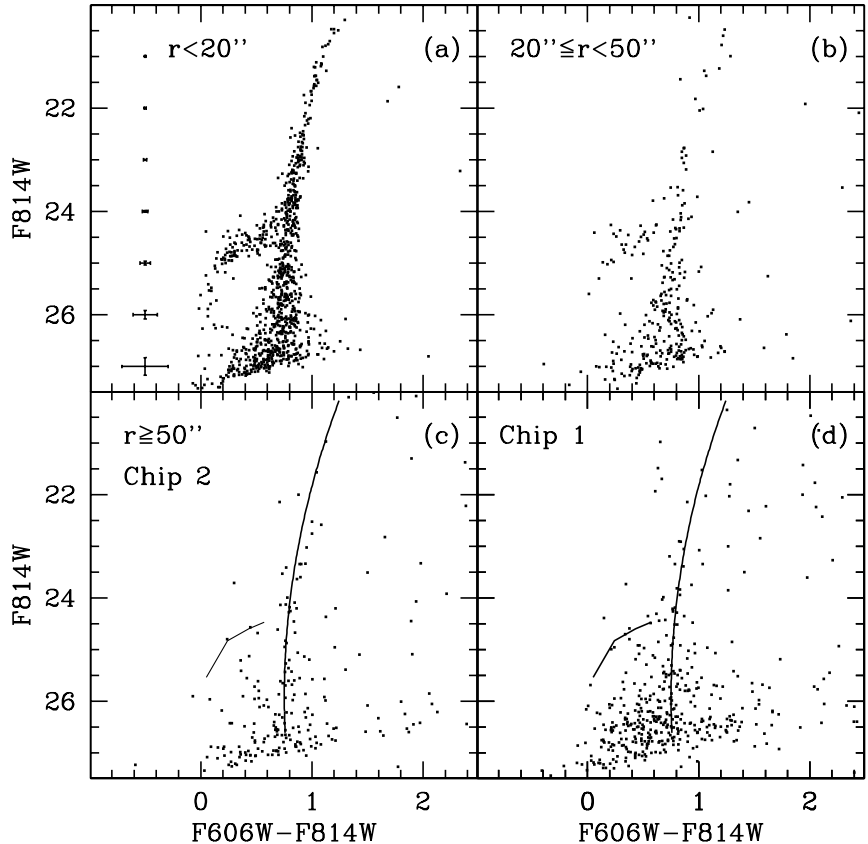}
\caption{Color Magnitude Diagrams of different radial annuli centered on the
center of the cluster (panels $a$, $b$, $c$,and $c$). The considered
annuli sample the following fraction of the area of Chip 2, approximately: 
($a$) 6\%, ($b$) 33\%, and ($c$) 60\%, respectively. The fiducial ridge lines of
the cluster are superposed to the CMD in panel $c$.
Panel $d$: CMD of all the stars in Chip 1 (sampling the field population) with
the cluster fiducial ridge lines superposed (solid lines).} 
\label{fig2}
\end{figure*}

\begin{figure*}[t]
\epsscale{.83}
\plotone{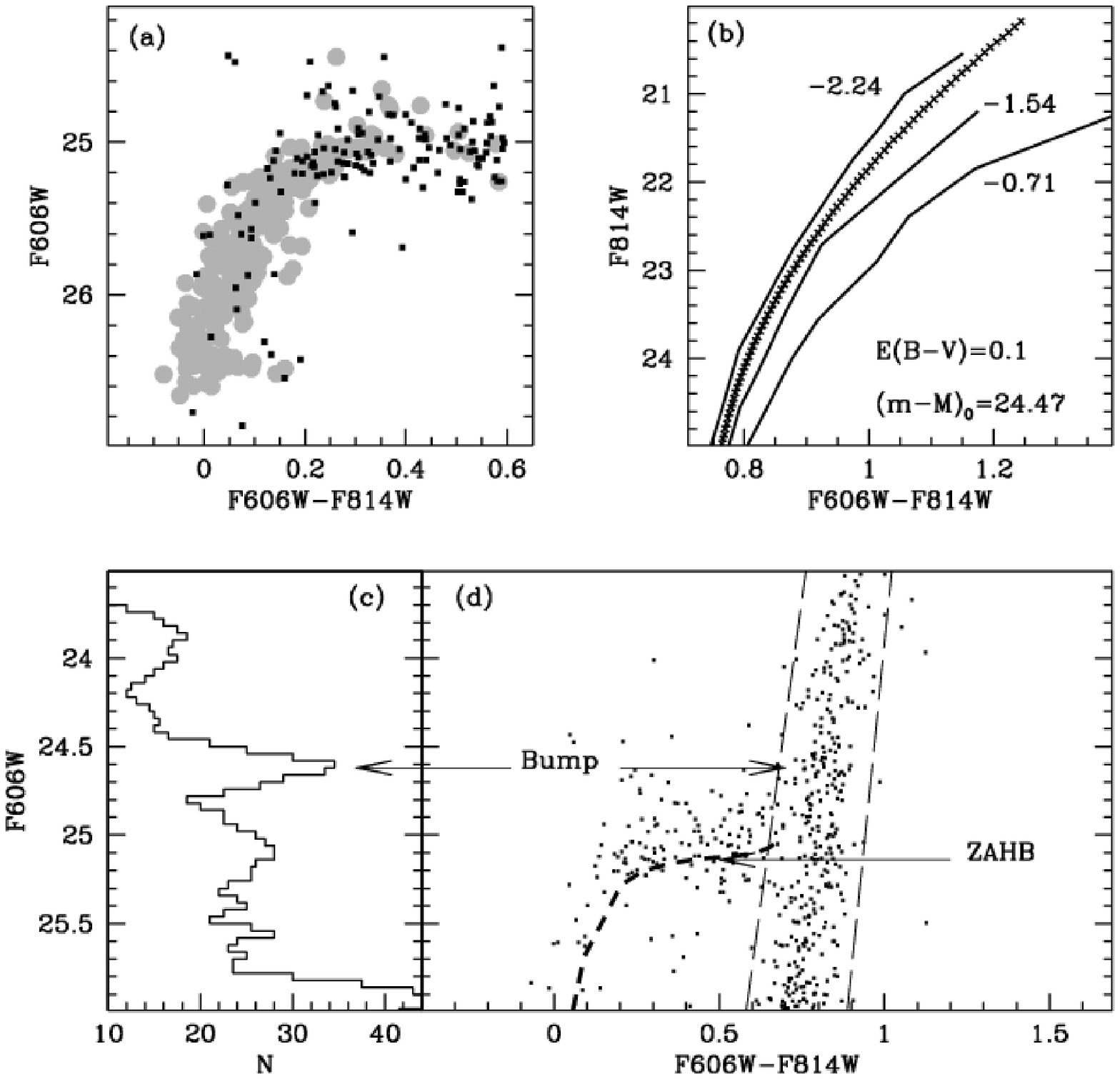}
\caption{Panel $a$: comparison between the (blue) Horizontal Branches 
of B514 (dark dots) and M92 (grey circles), obtained assuming E(B-V)=0.1 and 
$(m-M)_0=24.47$ for B514, similar to G05. 
Panel $b$: comparison between the RGB ridge line of B514 (crosses) an
those of template Galactic clusters (solid lines) from  
\citet{brown}. From the left to the
right the fiducials are M92, NGC~6752 and 47~Tuc. 
The corresponding metallicities are labeled on the plot. 
Panel $c$: luminosity function of the RGB of B514
presented as a smoothed histogram with step of 0.04 mag and bin of 0.2 mag.
The arrow indicates the position of the RGB bump. 
Panel $d$: CMD of B514 with the
position of the RGB Bump and the ZAHB indicated by arrows. The long-dashed line
is the ZAHB model for $Z=0.0003$ and $Y=0.245$ by \citet{basti} in the ACS
VEGAMAG photometric system, corrected for the assumed reddening and distance
modulus of B514. 
The dashed lines mark the region of CMD which was used to create the 
luminosity function in panel $c$.
}
\label{fig3}
\end{figure*}

In Fig.\ref{fig2} we plot the Color Magnitude Diagrams (CMD) of different radial annuli
around the cluster center in Chip 2 (panels $a$, $b$ and $c$) and of all  the
stars detected in Chip 1 (panel $d$). Most of the cluster stars are obviously
in the innermost region, hence the diagram of panel $a$ shows more clearly the
main evolutionary sequences. The CMD is dominated by an extended and very steep
Red Giant Branch (RGB) going from $F814W\sim 20.5$ and $F606W-F814W\simeq 1.1$
down to the limiting magnitude level at $F606W-F814W\simeq 0.9$. A small
overdensity corresponding to the RGB Bump (see below) is discernible at
$F814W\simeq 23.8$. The Horizontal Branch (HB) appears quite extended in color.
A sloped band of stars crossing the sequence at $F814W\simeq 24.6$ and 
$F606W-F814W\simeq 0.4$ has the typical appearance of a population of RR Lyrae 
stars sampled at random phase, strongly suggesting the presence of a
significant number of such variables (see also Fig.\ref{fig3}a, below).  The HB is
clearly well populated to the blue of the instability strip with a possible
hint to the presence of Extreme HB stars at $F814W>25.0$ and  $F606W-F814W\sim
0.0$. The mere presence of Blue HB stars and RR Lyrae indicates that the
cluster is probably older then 10 Gyr. The possible presence of HB stars
lying to the red of the instability strip needs a more detailed analysis to be
considered in the present context. This, as well as other issues, is demanded to
a future, more thorough contribution (Galleti et al., in preparation, hereafter
Pap II).

The diagram in panel $d$ shows that the stellar density in these extreme regions of 
M31 is so low  that the field of view of a single ACS/WFI
is insufficient to sample the evolved population of the M31 halo, hence no clear
RGB and/or HB sequence can be identified and the degree of contamination of the
cluster CMD should be negligible, at least for $F814W\le 26.0$. 
As a reference, we have overplotted in panel $d$ the ridge lines of the RGB and
HB of B514, derived as in \citet[][hereafter F99]{f99}.
Note that the
sky area sampled by the CMD of panels $a,b$ and $c$ is $\sim$6\%, $\sim$33\%,
and $\sim$60\%, respectively, of the area of Chip 1 (CMDs of panel $d$).


Panel $b$  of Fig.\ref{fig2} shows that cluster RGB and HB are clearly visible
out to a radius of $>20\arcsec$, corresponding to $\simeq 76$ pc at the
distance of M31 \citep[assuming $D=783$ kpc, from][]{mcc}.   A value of
$r=50\arcsec$ is the radius of the largest circle centered on the cluster that
is completely enclosed within Chip 2. In panel $c$ of Fig.\ref{fig2} we plot
Chip 2 stars having  $r\ge 50\arcsec$: the CMD is too sparsely populated here
to establish if cluster stars are found this far from the cluster center. While
the detailed analysis of the cluster profile will  be presented in Pap II we
already anticipate that a break in the profile is clearly detected around the
tidal radius of the cluster (at $r\sim 20\arcsec$), suggesting the  presence 
of extra-tidal stars \citep{katy}.

\subsection{Distance, metallicity and the RGB bump}

In Fig.\ref{fig3} we present a series of diagrams illustrating our assumptions on 
distance and reddening, our estimates of the metallicity of B514 and the 
detection of the RGB bump of the cluster:


{\bf Panel $a$:} We assume $(m-M)_0=24.47$, after \citet{mcc}, and $E(B-V)=0.1$,
similar to G05. To check the consistency of these assumptions we compare the BHB of
B514 with that of the metal-poor Galactic globular M92 (NGC 6341).
The ACS data for
M92 are from the set presented by \citet{brown}; here we have reduced and
calibrated the shortest exposures in a fully homogeneous way to obtain a
comparison with B514 as self-consistent as possible. The reddening and distance
modulus of M92 are taken from F99 (column 5 and 8, respectively, of their
Tab.~2). The HB morphology of M92 is sufficiently similar to the B514 one to
provide a wide overlap in the portion of the BHB that is more suitable to be
used as a standard candle, e.g. in the range 
$0.2\le F606W-F814W\le 0.35$, where the sequence is nearly horizontal.
The diagram clearly show that our assumed distance and reddening provide a very
good match between the BHB of the two clusters, fully supporting our choices. 

{\bf Panel $b$:} The RGB ridge line of B514 (already presented in Fig.~2, 
above) is compared with the RGB fiducials of the Galactic clusters M92,
NGC~6752, and 47~Tucanae, taken from \citet{brown}, corrected for our assumed
distance and reddening. The fiducial lines are
labeled with their [Fe/H] value in the \citet{zinn} scale, from F99.
Reddening and distance moduli of the templates are taken from F99. The RGB of
B514 is nicely enclosed between the fiducials at $[Fe/H]=-2.24$ and 
$[Fe/H]=-1.54$, in good agreement with the spectroscopic estimate of 
$[Fe/H]=-1.8$ provided by G05. 

{\bf Panel $c$:} The RGB bump is a well known feature in the luminosity
function of the RGB of globular clusters (see F99 for details and references).
The bump is clearly identified here as a peak in the LF of the RGB of B514, at
$F606W=24.62\pm 0.05$.
RGB stars were selected in the CMD using a color-magnitude window that follows
the RGB ridge line, excluding HB stars and field stars. 
A RGB bump brighter than the HB level is typical of metal
poor globular clusters.

{\bf Panel $d$:} A theoretical Zero Age HB sequence (with $Z=3\times10^{-4}$ and
Y=0.245) is superposed to the CMD of B514. The sequence is taken from the set by
\citet{basti}, transformed to the ACS system by \citet{bedin}, and is corrected
for the reddening and distance modulus assumed above. As expected, the
theoretical ZAHB lies at the faint end of the observed HB. We take the ZAHB
luminosity at $F606W-F814W=0.4$ as the fiducial ZAHB level,
$F606W_{HB}=25.14$. The magnitude difference between the bump and the HB is
$\Delta F606W_{HB}^{bump}=-0.62 \pm 0.1$. 
We transformed the F606W magnitudes of
the Bump and the ZAHB into Johnson-Cousins V magnitudes by using the relation 

$$V=F606W+0.205C^2+0.104C+0.037 ~~~~(1)$$

where $C=(F606W-F814W)$, that we have obtained from more than 400 stars in
common between our ACS photometry of M92 and the set of secondary photometric 
standards in M92 by \citet{stet}. 
From the obtained values, $V_{bump}=24.89$ and $V_{HB}=25.25$, we compute 
$\Delta V_{HB}^{bump}=-0.36\pm 0.12$ mag, that can be used as input for
the relations by F99 between 
$\Delta V_{HB}^{bump}$ and $[Fe/H]$, calibrated on Galactic globulars, to 
obtain an independent estimate of the metallicity from the Bump. 
From Eq.~6.2 of F99 we obtain $[Fe/H]=-1.8 \pm 0.15$, further supporting our 
previous estimates. 

We obtain also a preliminary estimate of the integrated V magnitude of the
cluster, by simple aperture photometry out to $r\simeq 14\arcsec$ from the 
cluster center. This aperture includes the majority of the cluster light but,
obviously, since the tenuous cluster halo extends much beyond this limit, the
derived integrated luminosity should be considered as a lower limit.
We obtain
$F606W\simeq 15.5$ mag and $F606W-F814W\simeq 0.8$, that are converted to 
$V\simeq 15.7$ mag according to Eq.~1.
Correcting for distance and extinction we find an absolute V magnitude 
$M_V\la -9.1$, corresponding to a luminosity of $L\ga 3.5\times 10^5 L_{\odot}$.

All the above analysis confirms that B514 is a genuine, very bright, old and
metal-poor globular cluster orbiting at a large distance from the center of M31.
The case of B514 demonstrates that the search for remote M31 clusters is a
promising line of research, that may finally lead to the assembly of a
sufficiently large sample of clusters to allow a deeper insight into the yet
poorly explored outskirts of our neighbour spiral galaxy.

\acknowledgments

We acknowledge the financial support to this research by Agenzia Spaziale
Italiana (ASI) and the Italian Ministero dell'Universit\`a e della Ricerca
under grant INAF/PRIN05 1.06.08.03.
Part of the data analysis has been carried on with software developed by P.
Montegriffo at INAF - Bologna Observatory.

\end{document}